# Home monitoring for frailty detection through sound and speaker diarization analysis

Yannis Tevissen[2,3], Dan Istrate[1], Vincent Zalc[1], Jérôme Boudy[2], Gérard Chollet[2],
Frédéric Petitpont[3], Sami Boutamine[1]

[1]Université de technologie de Compiègne, CNRS, Biomécanique et Bioingénierie, Centre de Recherche Royallieu – CS 60319 - 60203 Compiègne Cedex, France
[2]SAMOVAR, Télécom SudParis, Institut Polytechnique de Paris, 911200 Palaiseau, France
[3]Newsbridge, 92100 Boulogne-Billancourt, France
yannis.tevissen@telecom-sudparis.eu

*Abstract* – *As the French, European and worldwide populations are aging, there is a strong interest for new systems that guarantee a reliable and privacy preserving home monitoring for frailty prevention. This work is a part of a global environmental audio analysis system which aims to help identification of Activities of Daily Life (ADL) through human and everyday life sounds recognition, speech presence and number of speakers detection. The focus is made on the number of speakers detection. In this article, we present how recent advances in sound processing and speaker diarization can improve the existing embedded systems. We study the performances of two new methods and discuss the benefits of DNN based approaches which improve performances by about 100%.*

*Keywords: Speaker diarization, home monitoring, prevention, embedded system.*

## I. Introduction

According to the National Institute of Statistics and Economic Studies [1] on January 1, 2020, the aging of the French population continues to increase. The population of people aged 65 and over increased from 19.7% in 2018 to 20.5% in 2020. This aging has implications for most spheres of society and the economy.

As life expectancy increases, older people who continue to live at home are increasingly isolated. The older they get, the more likely they are to lose friends of their generation. Several factors such as the desire to remain independent, the remoteness of families both geographically and in terms of the time they can devote, increase isolation. Thus, alone at home, it can quickly become dangerous, especially due to functional limitations in some cases and lead to domestic accidents or falls.

Falling is the leading cause of death for people over the age of 65, resulting in approximately 9000 deaths each year [1], which makes it necessary to use technological solutions to ensure the safety and maintenance of elderly people in their homes in order to enable them to continue to live in complete assurance and autonomy. This means that the living conditions of the elderly person can be improved, their family and social ties preserved and that they can enjoy maximum independence. There is a range of solutions from new technologies, facilitating the home support of older people, such as smart sensors [2], [3], [4] allowing the identification of distress situations by detecting falls and thus reducing the physical and psychological consequences (slowing the effects of inactivity of older people).

Currently, the systems try to identify the elderly frailty in order to try to prevent fall. Using different environmental sensor, the activity and habits of elderly are identified, and all changes can indicate a possible frailty. Classical system uses infrared sensors to detect movement and infer the activities. The infrared sensors have the advantage of simplicity and lower cost, but the information is reduced. A solution is to add other sensors without entering the privacy of the user.

The solution we propose is to use environmental sound analysis to detect environmental sound but also to count the number of speakers.

## II. Audio environment analysis

The environmental sound is a rich information source and less intrusive that image through video camera. The environmental sound is composed by speech and everyday life sounds. From speech we can extract the pronounced sentences and information about speakers; as from sound we can recognize human sounds (cough, sneeze, screams, yawn…), sound produced by human actions (dishes, object falls, glass breaking …) and other sounds (ventilator, water, vacuum cleaner…).

The interest to recognize everyday life sounds is no longer to show. Continuous speech recognition is too intrusive but recognizing only some distress expressions ("Help!, Help-me!…") using keyword spotting systems is complementary to the sound. Applying a speaker diarization system allow to know if the elderly people receive visit (social link), has received the meals or the cleaning person.

The everyday life sounds analysis system based on i-vectors was described in [5], [7]. A first number of speaker detection system was already proposed [5].

## III. PREVIOUS EMBEDDED NUMBER OF SPEAKER DETECTION ON-LINE SYSTEM

The previous proposed system detecting online the number of speaker was integrated in the general architecture of the system allowing also the everyday life sound recognition (see Figure 1).

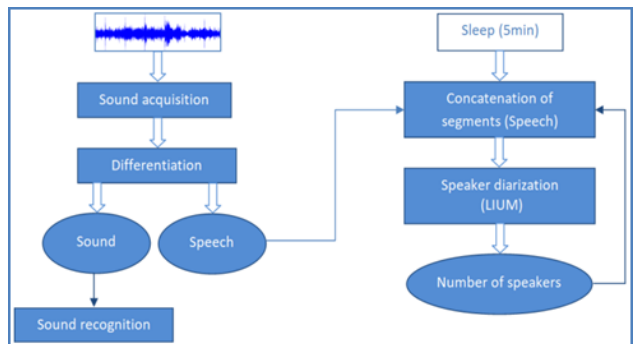

Figure 1. Sound and number of speaker recognition system

The system was designed to carry out an information about the number of speaker each 3 minutes (one minute if continuous speech) being able to turn on an embedded system.

For the hardware, the choice was made to use a Raspberry PI 4 and a Jabra 510 microphone. The software library was the LIUM speaker diarization system.

The speech identified after sound event detection step and differentiation sound/speech, is concatenated in order to obtain 3 minutes (this duration correspond to minimum duration needed in order to have acceptable performances). Each time that 3 minutes of speech are available the diarization system is launched and a number of speakers is estimated. A post-processing is added in order to eliminate the proposed speakers that has less than 10 seconds of duration. In order to have more often a number of speaker estimation, two minutes of speech from the current analysis are kept and only a new one minute is needed.

This system is integrated with sound analysis and has the advantage to make all the processing on the embedded device without speech transmission in the cloud. The system performances (see in the next section) are fair, and the estimated number of speakers doesn't consider the previous speakers.

For all that reasons, a new approach is described in the next sections.

## IV. PROPOSED SPEAKER DIARIZATION APPROACH

Recent advances on speaker diarization have shown particularly significant performance and robustness improvements [8], especially with the rise of deep learning based methods [9].

We propose here to use another algorithm for speaker diarization that have proven to be robust in some of the most complex scenari. This solution uses several voice activity detection (VAD) systems and merges them into one segmentation results by choosing for every 250 ms window the VAD system with the lowest entropy. The performance and robustness of this system have been demonstrated during the VoxSRC 2022 Challenge [10] and the implementation details are shown in [11].

The system we used here has two main differences with the system mentioned before.

First, it uses three different VAD systems: a C-RNN classifier based on a weakly supervised sound event detection scheme [12], the C-RNN from the pyannote toolkit [13], and the CRDNN from the speechbrain toolkit [14].

Secondly, instead of computing simple entropies as in [11], we compute for each classifier an normalized entropy so that the mean entropy of all systems is equal to 0.5. Then we compare at each 250 ms window the three entropies and choose the result from the classifier with the lowest one.

Once the VAD is computed we use a Bayesian HMM clustering method initialized with a standard agglomerative hierarchical clustering [16].

**Online capabilities**

As the goal of this system is to monitor elderly people at home, there is a strong need for near real-time results. To achieve this with the system mentioned before, we extract every 3 minutes an audio recording of the monitored room and we analyze it as a prerecorded file so our maximum delay time for decision making is around 4 minutes.

Knowing that every minute can be crucial in such contexts we also choose to study the performance trade off of using a completely online system such as the one described in [Coria et al. 2021]. As for the experimented we set its internal latency to 500 ms.

**Data used and evaluation method**

The proposed multi-stream voice activity detection system is evaluated on short conversations recorded with 1 to 4 active speakers. Each conversation last between 3 to 5 minutes and are performed by both male and female speakers.

For each recording, we perform speaker diarization and extract the number of different speakers detected. We compare these results with the ground truth to determine if the recording has been correctly labeled by the speaker diarization system.

Additionally, we also compute the Diarization Fairness Rate (DFR) described, for a set of recordings with only one active speaker, in [17] as follows:

$$DFR = \frac{Number\ of\ recordings\ with\ 1\ speaker\ detected}{Total\ number\ of\ recordings}$$

For each metric, we compare the results obtained by the MSVAD Diarization with the work previously done with the LIUM_SpkDiarization system described in [7] and with a state-of-the-art online diarization system.

## V. RESULTS

TABLE I. PERFORMANCES OF SPEAKER DIARIZATION SYSTEMS FOR NUMBER OF SPEAKER COUNT. CONFIDENCE INTERVALS AT 99% ARE ALSO REPORTED

| Method | Percentage of correctly labeled recording |
|---|---|
| LIUM_SpkDiarization | (32.5 ± 2.4) % |
| Diart (online) | (57.5 ± 3.1) % |
| MSVAD Diarization | **(77.5 ± 3.6) %** |

The results clearly demonstrate a significant performance improvement when using the recent speaker diarization systems. The online solution, even if it has the advantage of running continuously and producing results in a streaming way, performs achieve 20 % less correct detection than the MSVAD Diarization. For the latter we would need to perform 3 min splits similar as what was previously done in order to keep the continuity of the results on the embedded device.

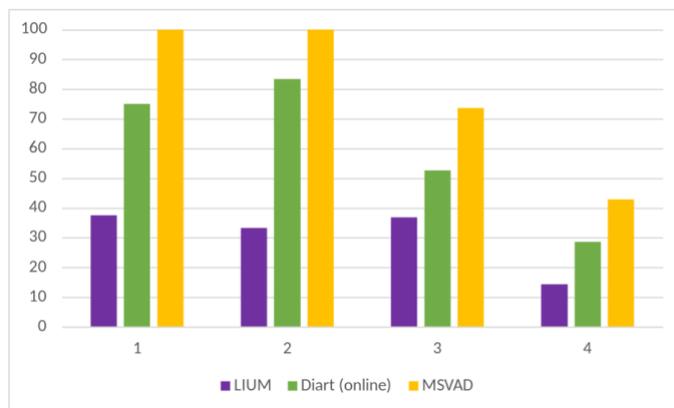

Figure 2. Detailed results of the percentage of correct speaker count for each system depending on the number of active speakers in the recording.

As expected, as we add speakers, the performances of the system decrease.

We also report the mean relative distance between the speaker count of the systems and the ground truth.

TABLE II. RELATIVE DISTANCE BETWEEN THE STUDIED SYSTEMS AND THE GROUND TRUTH NUMBER OF ACTIVE SPEAKERS.

| Method | Relative distance with ground truth labels |
|---|---|
| LIUM_SpkDiarization | 0.9 speakers |
| Diart (online) | 0.3 speakers |
| MSVAD Diarization | **0.1 speakers** |

TABLE III. FAIRNESS PERFORMANCES OF THE TWO SPEAKER DIARIZATION SYSTEM BASED ON RECORDINGS WITH ONE ACTIVE SPEAKER.

| Method | Diarization Fairness Rate |
|---|---|
| LIUM_SpkDiarization | 37.5 % |
| Diart (online) | 75.0 % |
| MSVAD Diarization | **100.0 %** |

These results show that, for recordings with only one active speaker, the MSVAD-based diarization never detects more than one speaker. Although this result must be put in perspective with the size of the database used, this is very encouraging for our home monitoring applications where we need to determine if the elderly person is alone or not.

## VI. CONCLUSIONS AND PERSPECTIVES

The results presented clearly show the benefits of using more recent DNN-based methods of speaker diarization in a home monitoring scenario as we improved our performances by more than 100%.

In this paper we demonstrated that both Diart and the MSVAD speaker diarization have advantages. To further choose between the two systems we will need to study their response time and to test them in real conditions.

As these methods tends to be more energy-consuming and we would need to assess the possibility to run them on small IOT devices such as a Raspberry Pi with an ARM CPU and a lower energy consumption. In a future work we will evaluate the possibility of system integration in the sound environmental system on embedded card.

## ACKNOWLEDGMENTS

This research work was funded by the FUI COCAPS project and the company Newsbridge.